# Tailoring energy landscape of graphene nanostructures on graphene and realizing atomically precise graphene origami using tilt grain boundaries

Yi-Wen Liu[§], Chen-Yue Hao[§], and Lin He[†]

Center for Advanced Quantum Studies, Department of Physics, Beijing Normal University, Beijing, 100875, People's Republic of China

[§]These authors contributed equally to this work.

[†]Correspondence and requests for materials should be addressed to L.H. (e-mail: helin@bnu.edu.cn).

**In two-dimensional van der Waals (vdWs) materials, the relative twist angle between adjacent layers not only controls their electronic properties, but also determines their stacking energy. This effect makes it much easier to realize energetically favorable configurations of the vdWs materials, for example, Bernal-stacked structure of bilayer graphene. Here we demonstrate that we can controllably tailor adhesive energy landscape of graphene nanostructures on graphene and stabilize the system with tunable twist angle by using a one-dimensional tilt grain boundary (GB). The area ratio with different stacking orders separated by tilt GB is continuously tuned, which provides a new degree of freedom to tailor the energy landscape of the system. Due to the different stacking orders separated by the tilt GB, we can repeatedly fold and unfold the graphene nanostructure exactly along the one-dimensional boundary, demonstrating the ability to realize atomically precise graphene origami.**

In van der Waals (vdWs) layered materials, the physical properties crucially depend on the relative twist angle between adjacent layers (*1-5*). The most reputed example is bilayer graphene in which we can obtain two-dimensional quasicrystals (*6-8*), magic-angle twisted bilayer graphene with flat bands (*9-11*), minimally twisted bilayer graphene with triangular network of chiral one-dimensional states (*12-15*), and so on, by simply varying the relative twist angle. The structures and the corresponding stacking energy in the bilayer graphene also depend on the relative twist angle. The bilayer graphene with stacking misorientation is energetically unstable and the zero-twist-angle stacking order, *i.e.*, the Bernal-stacked bilayer, is the most energetically favorable structure (*16-19*). Consequently, thermal fluctuations can lead to rotation between the adjacent vdWs layers to the energetically favorable structure (*20-26*). Therefore, stabilizing the twisted bilayer graphene and achieving precise angle control are of equal importance in the development of twistronics.

In this work, we demonstrate the ability to stabilize the stacking misorientation of graphene nanostructures on graphene by introducing a one-dimensional tilt grain boundary (GB), which is a line defect connecting two graphene grains with a relative rotated angle $\phi$ (*27-31*), in the supporting graphene. In our experiment, the area ratio with different stacking orders, as separated by the tilt GB, of the system is continuously tuned, which helps us to stabilize the graphene nanostructure with tunable twist angle on graphene. Because the different stacking orders separated by the tilt GB, we can repeatedly fold and unfold the graphene nanostructure exactly along the one-dimensional tilt GB. Our result demonstrates a new route to tailor adhesive energy landscape of bilayer graphene and provides a new method to realize atomically precise, custom-design graphene origami.

For a graphene nanostructure on a continuous graphene sheet, the relative twist angle between the adjacent layers plays a vital role in determining the resulting structure, as shown in Fig. 1A. Figure 1B shows the calculated adhesive energy of a graphene nanostructure with 3800 carbon atoms on a continuous graphene sheet, according to the theoretical result in ref. 18 (see supplemental material for detials, ref. *32*). The graphene nanostructure on graphene is initially at AB (or Bernal) stacking, defined as zero twist angle, and the adhesive energy depends sensitively on the relative twist angle. The energy maxima for the twist angles of 20 ° and 40 ° are set to zero (*18*). Obviously, the AB-stacked configuration is the most stable stacking order and the 30 ° configuration is a metastable stacking order. The configurations with other relative twist angles are not thermally stable configurations for the graphene nanostructure on graphene. Due to the structural symmetry of graphene, the adhesive energy as a function of twist angle exhibits a period of 60 °. By introducing a one-dimensional tilt GB in the supporting graphene sheet, as schematically shown in Fig. 1A, we can controllably tune area ratio $R_L$ of the graphene nanostructure between the left side and the right side separated by the GB. This introduces a new degree of freedom, *i.e.*, the $R_L$, to modulate the adhesive energy landscape of the graphene nanostructure on graphene. Figure 1C shows the calculated energy landscape of the graphene nanostructure on a supporting graphene with a $\phi = 21$ ° tilt GB. Now, the total energy of systems depends on both the twist angle and the area ratio $R_L$ separated by the GB. For the cases with $R_L = 100\%$ and 0, there is no GB in the supporting graphene. According to Fig. 1C, there are always one stable state and one metastable state, however, the corresponding angles of the stable state and the metastable state are not fixed but depending on the area ratio $R_L$. Figure 1D shows representative results with the area ratio $R_L = 25\%$, 50% and 75%, which exhibit quite distinct features comparing to that with $R_L = 100\%$ (or 0). For the $\phi = 21$ ° tilt GB, the stable state and the metastable state can be switched by changing the area ratio $R_L$. Moreover, the twist angle of two regions separated by the GB can be continuously tuned from 0 ° to 9 ° and from 21 ° to 30 °, respectively.

To explore above effects experimentally, highly oriented pyrolytic graphite (HOPG) and multilayer graphene on Ni foil synthesized by chemical vapor deposition (Fig. S1) are used in our experiment. Previous studies have demonstrated that one-dimensional tilt GBs with different $\phi$ can be frequently observed in the two systems (*27-31*), as also observed in our scanning tunneling microscopy (STM) measurement (Fig. S2). The graphene nanostructures are obtained by etching the studied samples with hydrogen plasma etching method (Figs. S3-S4). To study the effects of the tilt GB on the stacking orders of the graphene nanosctructures on graphene, we use a STM tip to "move" an etched graphene nanosctructure to a selected tilt GB. Figures 2A-2D summarize two typical processes in our experiment and we can repeatedly move the graphene nanostructure on and off the tilt GB (Fig. S5). Our experiment indicates that the graphene nanostructure on the tilt GB is quite stable during the measurement and the graphene nanostructure becomes movable only when the STM tip is approached to manipulate it. To fully understand the experimental result, we carried out atomic-resolved STM measurements (Figs. S6-S7). The rotated angle of crystal orientation between the left and right sides of the tilt GB in the supporting graphene is about 21.1°±1.0°, and the relative twist angles $\theta$ of the graphene nanostructure to the left region of the tilt GB in Figs. 2A-2D are also measured. According to the area ratio $R_L$ and twist angle $\theta$ with the left side of the GB, we can obtain the detected initial and final states in Figs. 2A-2D on the energy landscape, as summarized in Fig. 2E. Obviously, the final configurations in Figs. 2B and 2D are in the valleys of the energy landscape, indicating that they are energetically favorable states, as observed in our experiment. The competition of the adhesive energy between the two regions separated by the tilt GB helps to stabilize the configurations with stacking misorientations in the studied system.

When there is a GB in the supporting graphene, the energy barrier between the stable state and the metastable state depends on the area ratio and is much smaller than the cases with $R = 100\%$ and 0, as shown in Fig. 1. This indicates that it will be relatively easier to switch the configurations between the stable state and the metastable state. Recent experiments demonstrated the ability to tune strained structures of graphene by using local probing tip (*33-42*). In this work, we demonstrate that we can controllably switch the configurations between the stable states and the metastable states by using STM tip. Figures 3A-3C summarize three representative results obtained in our experiment. The initial and final configurations of the graphene nanostructure on the tilt GB are seized in the STM measurements and the corresponding states on the energy landscape are obtained, as shown in Fig. 3D, according to the area ratio and twist angle with the left side of the GB. From the energy landscape, the relationship among twist angle, area ratio and total energy can be vividly displayed. When the area ratio is close to 50%, the two regions separated by the tilt GB are stabilized in configurations with stacking misorientation. Increasing the difference of area between two regions, e.g. $R_L$ = 20% or 80%, the region with large area becomes aligned with the supporting graphene. Obviously, the changes of the configurations of the graphene nanostructure correspond to the switch between the stable states and the metastable states. The continuously tunable area ratio and twist angle of the system lowers the energy barrier between the stable state and the metastable state, which plays an important role in realizing the switch between them.

Due to the existence of the tilt GB, the two regions of the graphene nanostructure, as separated by the tilt GB, exhibit different stacking orders. Therefore, the vdWs forces between the two regions of the graphene nanostructure and the underlying graphene are different. Such a feature provides us unprecedented opportunity to fold the graphene nanostructure at position with atomically precision, as schematically shown in Fig. 4A. With decreasing the distance between the STM tip and the graphene nanostructure, we can lift and fold the graphene nanostructure at position of the tilt GB when tip-graphene vdWs force overcomes the force between the graphene nanostructure and the underlying graphene. Figure 4B shows a typical experimental result. The scanning bias is decreased from 1 V to 50 mV step by step with an interval of 50 mV. Then, the left side of the graphene nanostructure can be lifted and folded at position along the tilt GB under the scanning bias of 50 mV (see Fig. S8 for the folding and unfolding processes). The continuously tunable area ratio of the graphene nanostructure enables us to fold and unfold the graphene nanostructure at any selected position with atomically precision, as demonstrated in Figs. 4B-4D. Although fold and unfold graphene nanostructures along an arbitrarily chosen direction are realized very recently (*33*), fold and unfold the graphene nanostructures at atomically precise, custom-design position are still very big challenge in experiment. Our result provides a promising route to overcome this challenge with the help of the one-dimensional tilt GB.

In summary, through introducing a one-dimensional tilt GB in the supporting graphene, we provide a new route to tailor adhesive energy of the graphene nanostructures on graphene. The tunable area ratio with different stacking orders, as separated by the tilt GB, of the system helps us to stabilize the graphene nanostructure with tunable twist angle on graphene. Because the different stacking orders separated by the tilt GB, the graphene nanostructure can be folded and unfolded accurately along the GB. Our results provide a new route to realize atomically precision, custom-design origami of graphene, which can be applied in other two-dimensional materials.

**Acknowledgements**

This work was supported by the National Natural Science Foundation of China (Grant Nos. 11974050, 11674029). L.H. also acknowledges support from the National Program for Support of Top-notch Young Professionals, support from "the Fundamental Research Funds for the Central Universities", and support from "Chang Jiang Scholars Program".

**Author contributions**

Y.W.L performed the STM experiments. Y.W.L. and C.Y.H analyzed the data. L.H. conceived and provided advice on the experiment, analysis, and the theoretical calculation. L.H., C.Y.H and Y.W.L wrote the paper. All authors participated in the data discussion.

**Competing financial interests**

The authors declare no competing financial interests.

**Figures**

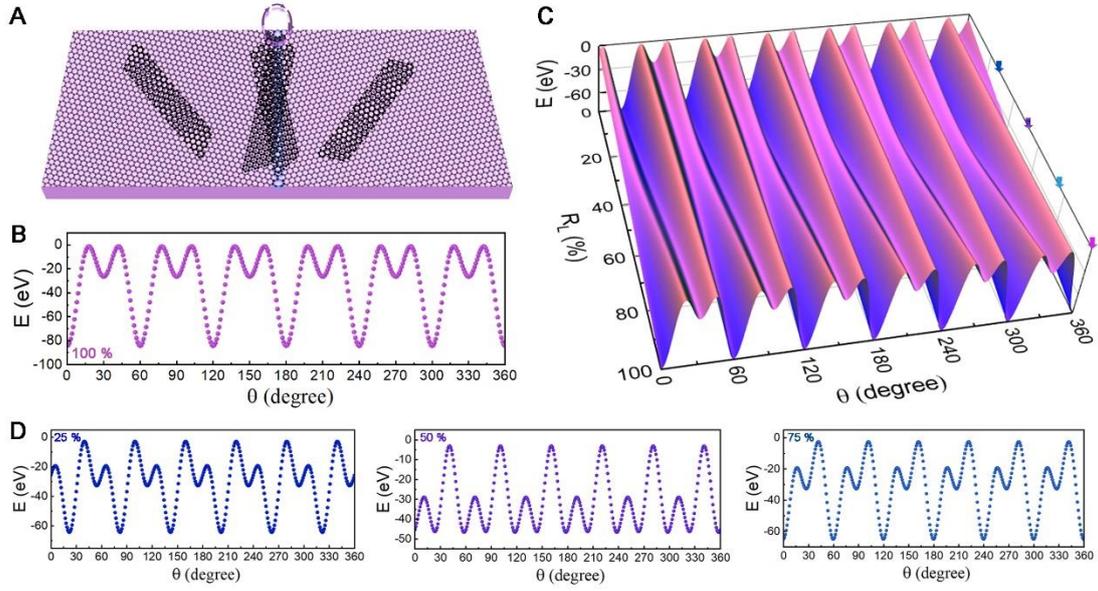

**Fig. 1. Energy landscape of graphene nanostructure on graphene with a tilt GB.**
**(A)** Schematic configurations of a graphene nanostructure on graphene with a tilt GB.
**(B)** The energy evolution of a graphene nanostructure on a single-crystal graphene with different relative twist angle. **(C)** The energy landscape of a graphene nanostructure on graphene with a tilt GB. The crystal orientation between left and right sides of the GB is assumed to be 21°. The parameter $\theta$ represents the twist angle of the graphene nanostructure with the underlying graphene on the left of the GB. The $R_L$ is defined as the area ratio between the region on the left of the GB and the whole nanostructure. **(D)** The energy evolution as a function of the twist angle for the cases that the area ratio is about 25%, 50% and 75 % respectively.

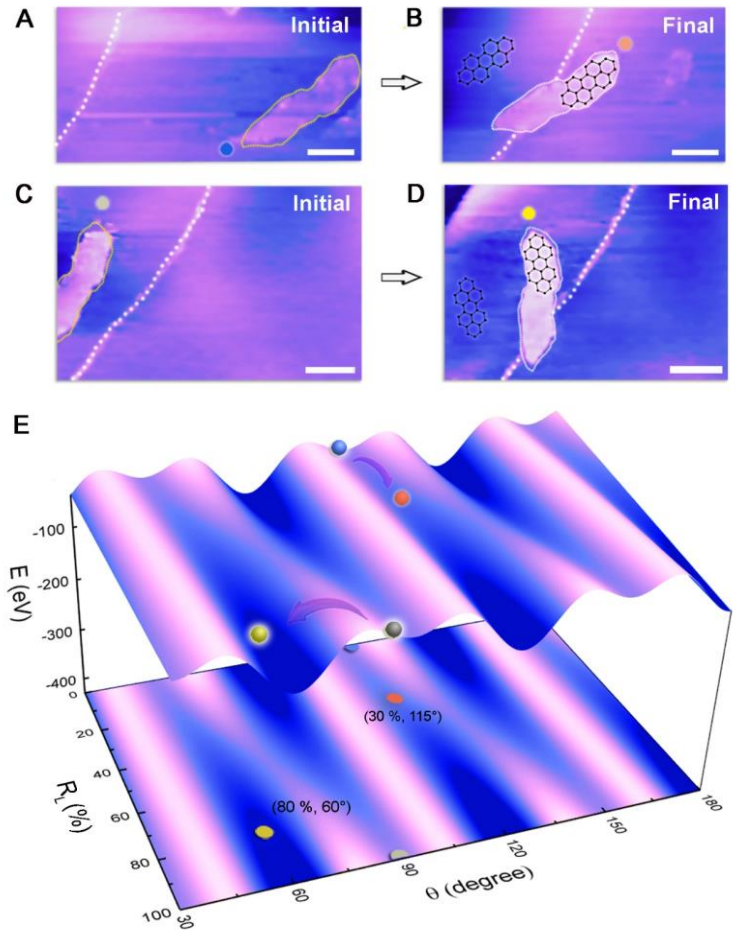

**Fig. 2. Stabilizing the graphene nanostructure by the tilt GB. (A)** STM image of an initial configuration showing that the graphene nanostructure is in the right side of the GB. The GB is highlighted with the white dot line. **(B)** STM image of a final configuration that the graphene nanostructure is "moved" to on top of the GB. The area ratio is measured as about 30% and the twist angle is measured as about 115°±1.0°. **(C)** STM image of an initial configuration showing that the graphene nanostructure is in the left side of the GB. **(D)** STM image of a final configuration that the graphene nanostructure is "moved" to on top of the GB with the area ratio about 80% and the twist angle about 60°±1.0°. Scale bar: 10 nm. **(E)** The corresponding states in energy landscape of the configurations in panels (A)-(D). The blue and grey dots indicate the initial states in (A) and (C) respectively. The yellow and orange dots show the final states in (B) and (D) respectively. The area ratio and the twist angle of these states are marked with ($R_L$, $\theta$).

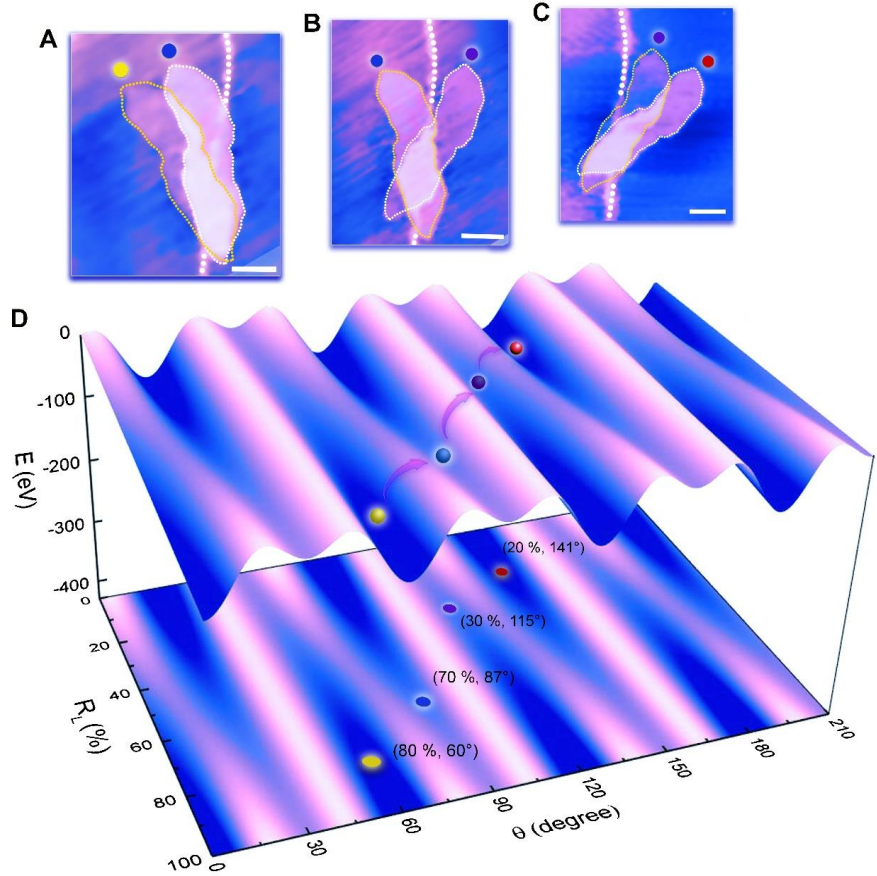

**Fig. 3. Tunable configurations of the graphene nanostructure on graphene with the tilt GB.** (**A**) The configurations of graphene nanostructure before (yellow dot) and after (blue dot) the first STM tip manipulation. (**B**) The STM image of the graphene nanostructure before (blue dot) and after (purple dot) the second STM operation. (**C**) The STM image of the graphene nanostructure before (purple dot) and after (red dot) the third STM manipulation. Scale bar: 8 nm. (**D**) The corresponding states in energy landscape of the configurations in panels (A)-(c). The corresponding states are identified with the colored dots with the area ratio and the twist angle ($R_L$, $\theta$).

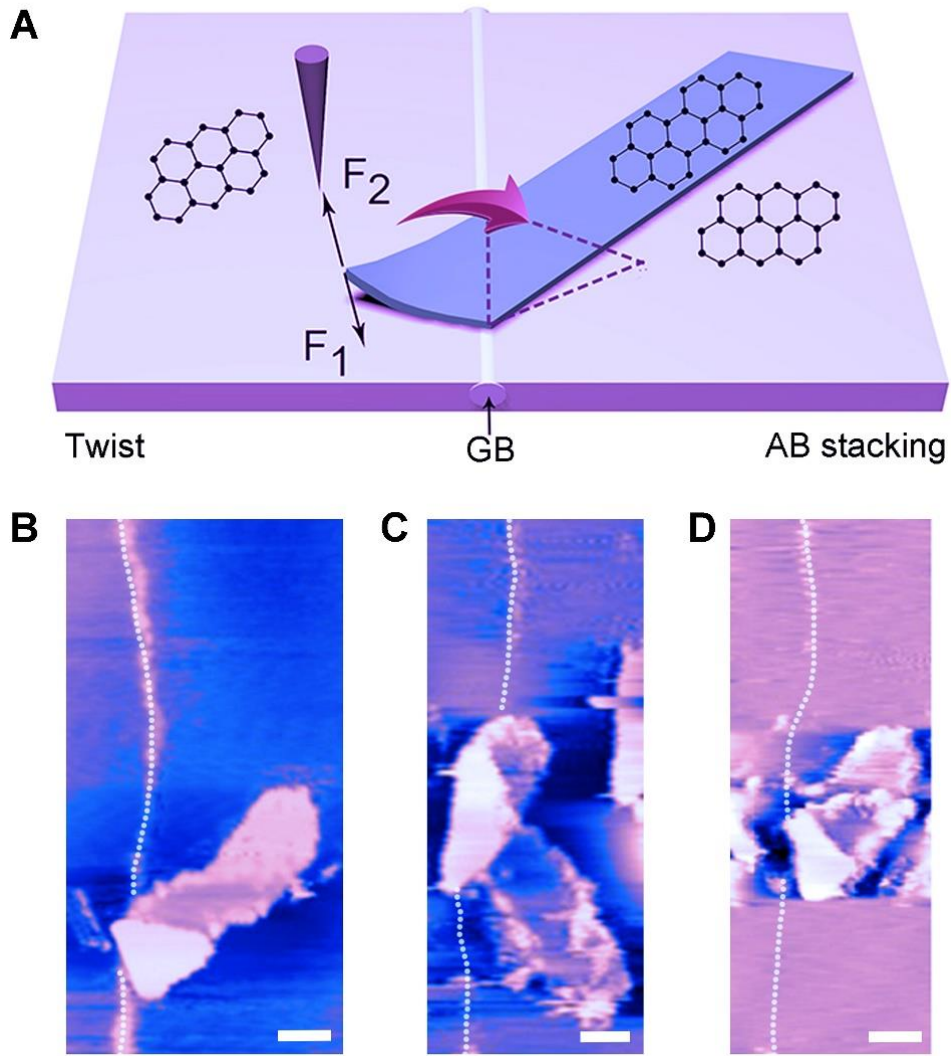

**Fig. 4. Atomically precise, custom-design origami using the tilt GB. (A)** Schematic figure shows the process of folding the graphene nanostructure along the tilt GB by using a STM tip. $F_1$ is the vdWs force between the graphene nanostructure and the underlying graphene. $F_2$ is the vdWs force between the STM tip and the graphene nanostructure. The purple dotted line is the final configuration of the graphene nanostructure after the folding. **(B-D)** STM images of the graphene nanostructure after the folding. The left side of the nanostructure with weaker interlayer interaction are folded to the right side along the one-dimensional tilt GB. Scale bar: 10 nm.